\documentclass[aps,prl]{revtex4}

\usepackage{graphicx}

\begin{document}

%Title of paper
\title{Balancing with Noise and Delay}

\author{Tadaaki Hosaka}
\affiliation{Department of Computational Intelligence and Systems Science, Tokyo Institute of Technology, Yokohama, Japan 226-8502}
\author{Toru Ohira} 
\affiliation{Sony Computer Science Laboratories, Inc., Tokyo, Japan 141-0022}
\author{Christian Luciani}
\affiliation{Departamento de F\'{\i}sica, Universidad Sim\'on Bol\'{\i}var, Caracas 1080-A, Venezuela}  
\author{Juan Luis Cabrera}
\affiliation{Centro de F\'{\i}sica, I.V.I.C., Caracas 1020-A, Venezuela} 
\author{John G. Milton}
\affiliation{Joint Science Department, Keck Science Building, The Claremont Colleges, Claremont, CA 91711, USA}

\begin{abstract}
Motivated by recent studies in human balance control, we study a delayed random walk with an
unstable fixed point.
It is observed that the random walker moves away from the unstable fixed point more slowly than
is observed in the absence of delay. 
It is shown that, for given a noise level, there exists an optimal 
delay to achieve the longest first passage time.
Our observations support recent demonstrations that noise has a beneficial role for balance
control and emphasize that predicitive strategies are not necessary to transiently control
balance. 

\vspace{1em}
\noindent
(Published in Progress of Theoretical Physics (Supplement) {\bf 161}, 314--319 (2006))
\end{abstract}

\date{\today}
\maketitle

\section{Introduction}

Feedback delay and noise pose challenging problems for the neural control of balance.
The time scales for the controlling motion in human postural sway \cite{Zatsiorsky-Duarte00}
and stick balancing at the fingertip \cite{cabrera-milton02,cabreraetal04} are much shorter
than the estimated latencies for the neural control mechanisms.
How is it possible to control noisy, unstable systems whose time constant is much shorter than 
the control time delay?  

Paradoxically, noise has a beneficial effect for balancing tasks 
\cite{cabrera-milton02,Priplata03}.
Recently it has been shown that the fluctuations in the controlled variable for stick balancing
at the fingertip, i.e. the vertical displacement angle $\theta$ , exhibit on--off intermittency \cite{cabrera-milton02}: 
a -1/2--power law is present in the power spectrum and a -3/2 power law occurs for the laminar phases.
Importantly, the time between $> 98\%$ of the successive controlling movements is less than the estimated neural
latency.
These phenomena can be reproduced by (\ref{juan_luis}), i.e. a damped inverted pendulum stabilized
by time--delayed feedback with a noisy gain, provided that the parameters are chosen to place the
system close to a stability boundary.
Thus the mechanism for obtaining fast control in the presence of time--delayed feedback is related, at least in
part, to the stochastic or chaotic forcing of important control parameters, e.g. gain, across a stability
boundary. 

What is the mechanism for the stabilizing effects of the state--dependent noise?
The stability boundary for (\ref{juan_luis}) separates a stable fixed point from an unstable fixed point \cite{cabrera-milton02}.
It has been suggested that the beneficial effects of on--off intermittency for stick balancing arise because
the fluctuations in $\theta$ resemble a random
walk for which the mean value of $\theta$ is approximately zero, i.e. the balanced position is statistically
stabilized \cite{cabrera-milton02}.
This interpretation suggests that the control parameters for stick balancing could be tuned to place the
system either directly on the stability boundary or perhaps to the stable side of the boundary; the proximity to the 
stability boundary depending on the intensity of the state--dependent noise.
Subsequently two experimental observations suggested that the system may actually be tuned slightly to the unstable
side of the stability boundary:
1) power laws identical to those observed for stick balancing are observed in a virtual stick balancing task that 
does not possess a stable fixed point \cite{cabreraetal04}; and 
2) the observed survival function for stick balancing, i.e. the probability that a stick is still balanced at 
time $t$, is well described by a second-order stochastic discrete equation that possesses only an unstable fixed 
point \cite{cabrera-A}.  
Taken together these observations lead to the surprising conclusion that an unstable fixed point can be stabilized 
in a dynamical system with retarded variables using state--dependent noise.
The obvious question is does such a mechanism exist?  

Here we study the first passage times of a delayed random walk in which the fixed point is unstable
or repulsive. 
We find that in a repulsive delayed random walk the first passage time is prolonged for certain values of the delay.
In other words, memory effects can transiently stabilize an unstable fixed point.
These observations challenge the hypothesis that the neural control of balance solely involves the use
of predictive control strategies \cite{mehta-schaal02}.

\section{Model}

From a mathematical point of view, the study of stochastic dynamical systems with retarded variables leads
naturally to formulations in terms of a stochastic delay-differential equation.
In the case of stick balancing at the fingertip, it was proposed 
that the dynamics of the deviation angle $\theta$ of the stick from the
upright position during the human stick balancing task can be described by \cite{cabrera-milton02}
\begin{equation}
m {d^2 \over dt^2} \theta(t) + \gamma {d \over dt} \theta(t)+ \kappa \theta(t) = (r_0 + \xi(t))\theta(t-\tau)
\label{juan_luis}
\end{equation}
where $m, \gamma, \kappa, r_0$ are parameters with $\xi(t)$ being the Gaussian white noise with strength $D$ as
follows,
\begin{eqnarray}
\langle\xi(t)\rangle &=& 0, \\
\langle\xi(t)\xi(t')\rangle &=& D \delta(t-t').
\end{eqnarray}
The time delay arises because of the neural latency in the initiation of the corrective movements.
Depending on the choice of parameters the upright position can be stable or unstable.
In order to reproduce the experimentally observed power laws it is necessary to chose the parameters close, or on, a 
stability boundary.

The study of the behavior of (\ref{juan_luis}) in the unstable regime is not straightforward.
Even today, the complexities involved in the study of stochastic delay equations in the stable
regime impose formidable barriers \cite{eurich96,kuchler,milton-mackey00}. 
In particular, the location of the stability boundary is uncertain if the intensity of either the
state--independent or state--dependent noise is unknown \cite{longtinetal90,mackey95}. 
An alternate approach which is amenable to analysis is to approach the problem from the perspective of
a delayed random walk \cite{ohira,ohira-sato,ohira-yamane} in which a walker takes a unit discrete step
along one dimensional axis and the position of the random walker at time step $t$ be given by $X(t)$.
The dynamics of the delayed random walk are governed by conditional probabilities which depend on the position of the walker
at a time $\tau$ in the past
\begin{eqnarray}
P(X(t+1)= X(t)+1|X(t-\tau)>0) & = & p \\
P(X(t+1)= X(t)+1|X(t-\tau)=0) & = & {1\over2} \\
P(X(t+1)= X(t)+1|X(t-\tau)<0) & =& 1-p,
\end{eqnarray}
where $0 < p < 1$ and $\tau$ is the delay. 
In this class of model the delay corresponds to the neural delay and the conditional probabilities are determined by 
the stability of the fixed point.
It should be noted that the fact that the conditional probabilities depend on the position of the walker they represent a
form of state--dependent noise.
Such models \cite{ohira-milton} have already provided some insights into the nature of the fluctuations observed in human 
postural sway \cite{collins-deluca}. 
 
For simplicity, set the fixed point at the origin, $X=0$.
With delay, the walker refers to its position in the past to decide on the bias, $p$, of the next step. 
%Previously we considered the case when $p < 0.5$, i.e. the case when
%the origin is attractive with no delay ($\tau=0$) \cite{ohira-milton,ohira,ohira-yamane}. 
Since the origin is repulsive we take $p > 0.5$. 

\section{Analysis and Simulation}

In contrast to the attractive case, as the walker escapes away from the origin we do not have a stationary 
probability distribution. Hence, we need to look at the other statistical aspects to
characterize its behavior. We focus here on the average first passage time and show
that delay induces a peculiar behavior. 

We consider the average first passage time $L$ to reach a certain position 
away from the origin, herein referred to as the limit point. 
In other words, we measure the average time for the walker starting from the origin to reach the limit 
point for the first time as we change parameters in the model. 

For the case of zero delay with the bias $p$, we can find an analytical result 
for this average first passage time $L$ to reach the limit point $\pm X^{*}$ as
\begin{equation}
L = 2 \left({q\over{q-p}} \right)
         \left({{1- \left({q \over p} \right)^{X^*}}\over{1-{q \over p}}} 
\right)
         +{X^* \over {p-q}}, \quad (p \neq 0.5),
\end{equation}
where we have set $q \equiv 1-p$. For the case of simple (symmetric) 
random walk with
$p=q=0.5$, this result reduces to an even simpler form.
\begin{equation}
L = (X^*)^2.
\end{equation}
 
For the case of non-zero delay, such analytical result is yet to be obtained 
and computer simulation is used. We considered an ensemble of 10000 
walkers. The initial condition is set so that the walker performs a normal 
random walk with no bias $p=0.5$ for the duration of $t=(-\tau,0)$. The 
walker's position at $t=0$ is
set as the origin $X=0$. The limit point is set at $\pm X^{*}$. We measure 
the number of steps for each walker to go from the origin to $\pm X^{*}$ and 
average them. We performed computer simulations for various bias  
$p$ and delay $\tau$. 

\begin{figure}[htbp]
\includegraphics[width=1.0\textwidth]{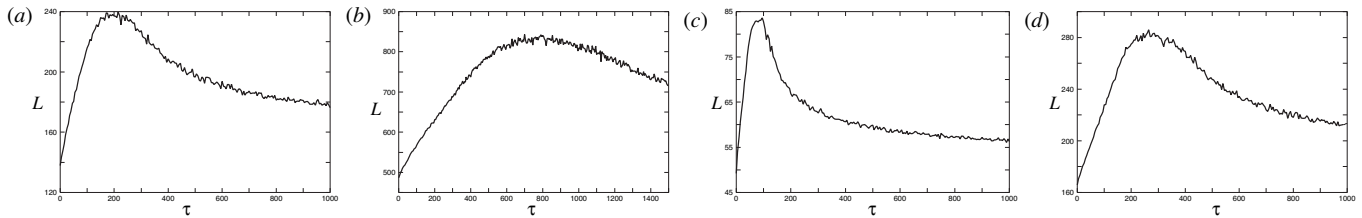}
\caption{Average first passage time $L$ as we change $\tau$. The value of 
parameters
$(p, {X^{*}})$ are (a) $(0.6, 30)$, (b) $(0.6, 100)$, (c) $(0.8, 30)$, and (d) $(0.8, 
100)$. }
\label{first_passage}
\end{figure}

Some sample results are shown in Figure \ref{first_passage}. 
The most notable features of these graphs are the peaks in the graph, indicating that the longest first passage 
time 
appears at certain optimal values of $\tau$ given bias $p$.
In other words, the walker is most stabilized around the origin with 
appropriate non-zero delay.
This is rather unexpected result contrary to the normal notion associated 
with effects of feedback delay,
where longer delay increasingly de-stabilize systems. Here, appropriate 
combination of bias and delay time is inducing more
stability.

In order to gain more insight into this phenomenon, we look for an 
approximate analytical expression, which is found to be given by the 
following expression 
\begin{equation}
L(\tau) =(1 + \alpha \tau_n {e^{ -\beta \tau_n }})L(\tau=0).
\end{equation}
Here $\alpha$ and $\beta$ are parameters and
${\tau_n}$ is a normalized delay given as follows.
\begin{equation}
{\tau_n}\equiv \tau { p-q \over {X^{*}} }.
\end{equation}
This normalization uses a characteristic time dividing the distance 
$X^{*}$ by an average velocity of the walker
 $p-q$. Hence ${\tau_n}$ is a non-dimensionalized parameter as well.

\begin{figure}[htbp]
\begin{center}
\includegraphics[width=0.3\textwidth]{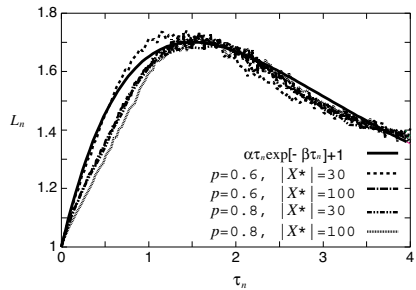}
\end{center}
\caption{Normalized average first passage time $L_n\equiv {L(\tau)\over L(\tau=0)}$ as we change 
normalized delay $\tau_n$. The parameter sets $(p, {X^{*}})$ plotted 
are $(0.6, 30)$,  $(0.6, 100)$, $(0.8, 30)$,  and $(0.8, 100)$ }
\label{delay_first_passage}
\end{figure}

Figure \ref{delay_first_passage} shows this analytical approximation and the result of computer 
simulation. We see that the curves for various  $(p, {X^{*}})$ overlaps quite 
well with the analytical curve with appropriately chosen parameters of 
$\alpha=1.27$ and $\beta=0.67$. We also notice that the peak height is 
approximately 1.7 times the average first passage time of zero delay case.

\section{Discussion}

A fundamental concept in the control of movement is that that nervous system calculates all of the
forces required to produce a voluntary movement prior to its execution.
Thus the control of movements on time scales faster than permitted by finite neural latencies is possible
because all of the necessary plans have been made beforehand.
However, even in well documented cases of predictive control such as stick balancing at the fingertip
\cite{mehta-schaal02}, the survival times for stick balancing obey statistics predicted for an unstable
fixed point with noise and delay \cite{cabrera-A,cabrera-B}.
Thus this theory is not without its problems \cite{ostry-feldman03}.
Moreover emphasis on predictive control strategies seems contrary to work on the development
of expertise suggesting that the ultimate goal of the nervous system is to minimize the need for active
predictive control (for a recent review see \cite{miltonetal04}).

Our study is the first to show that the interaction between state-dependent noise and a time delay can transiently
stabilize an unstable dynamical system. 
Obviously this form of stabilization does not involve predictive control, but depends on the relative values of the 
bias $p$ and delay $\tau$. 
Recent studies of the development of expertise in stick balancing have also shown that with increased skill
changes occur in both delay and the statistical properties of the controlling movements \cite{cabrera-B}.
Preliminary simulations indicate that the delayed random walk with repulsive origin reproduces the characteristics
of the survival times observed for stick balancing (C. Luciani, T. Hosaka, T. Ohira, J. G. Milton,
and J. L. Cabrera, unpublished observations). 
This observation is consistent with our suggestion that the control for stick balancing may be tuned to the unstable side
of the stability boundary.

However, the precise nature of this stabilizing mechanism is not yet clear.
The observations in Figures \ref{first_passage} and \ref{delay_first_passage} suggest that it could be related to 
a ``resonance--like'' phenomenon, such as that shown previously for delayed stochastic binary elements \cite{ohira-sato,tsimring}.
Another possibility is that it might be directly related to memory effects that arise through the influence
of the initial function on the subsequent evolution of the dynamical system.

The addition of noise can improve balance control in the elderly as measured from the fluctuations
recorded on a force platform \cite{Priplata03}.
However, it is not yet known to what extent these noise-induced changes in fluctuations in the center of 
pressure translate into a reduced risk of falling in the elderly \cite{moss-milton03}.
Could we do better if the effects of noise and delay on unstable dynamical systems were better understood?
One possibility, suggested by our results, is that balance could be improved if the appropriate level of fluctuations could be
introduced. 
There are various ways that this could be achieved. 
For example, as a test case, we have performed a preliminary experiment in which a subject is asked to shake an object with the other
hand at the same time he balances a stick on his fingertip. 
In this way, fluctuations are meditated through a subject's body to his
balancing fingertip. 
Though it takes some practices to find one's optimal way of generating a fluctuation, it is found that 
about $60\%$ of the subjects can better balance the stick with fluctuation \cite{ohira-hosaka05}. 
More experiments are required to explore such effects.

\subsubsection*{Acknowledgments}

JM acknowledges support of grants from the Kenan Foundation and the National
Institutes of Mental Health.
TH acknowledges support of the Japan Society for the Promotion of Science (JSPS) Research Fellowships for Young Scientists (Grant in Aid No. 164453).

\end{document}